\documentclass[11pt]{article}
\usepackage{moriond,epsfig}
\usepackage{floatflt}

\def\Journal#1#2#3#4{{#1} {\bf #2}, #3 (#4)}

\def\PLB{{\em Phys. Lett.}  B}
\def\PRL{\em Phys. Rev. Lett.}
\def\PRD{{\em Phys. Rev.} D}

\def\ra{\rightarrow}

\def\be{\begin{equation}}
\def\ee{\end{equation}}
\def\bea{\begin{eqnarray}}
\def\eea{\end{eqnarray}}

\def\Vts{$V_{ts}$}
\def\Vtd{$V_{td}$}

\def\Vub{$V_{ub}$}
\def\MVub{$|V_{ub}|$}
\def\Vcb{$V_{cb}$}
\def\MVcb{$|V_{cb}|$}

\def\btodlnu{$B\ra D\,\ell\,\nu$}
\def\btodstlnu{$B\ra D^*\,\ell\,\nu$}
\def\btoxhlnu{$B\ra X_H\,\ell\,\nu$}
\def\btodst{$B\ra D^*$}
\def\btosg{$b\ra s\,\gamma$}
\def\btoxclnu{$B\ra X_c\,\ell\,\nu$}
\def\btoc{$b\ra c$}
\def\btou{$b\ra u$}
\def\CP{$CP$}
\def\Lbar{$\bar{\Lambda}$}
\def\lone{$\lambda_1$}
\def\ltwo{$\lambda_2$}

\def\UpsIVS{$\Upsilon(4S)$}

\def\BBbar{$B\bar{B}$}

\def\btoclnu{$b\ra c\,\ell\,\nu$}
\def\Fdstwmin{$F_{D^*}(w_{\rm min})$}

\def\nGeV{{\rm GeV}}

\def\ninvfb{{\rm fb}^{-1}}
\def\nUnit#1#2{${#1}\,{#2}$}

\def\CLEOII{CLEO~II}
\def\CLEOIIV{CLEO~II.V}
\def\etal{{\it et al.}}
\begin{document}
\vspace*{4cm}
\title{$B$ MESON DECAYS: RECENT RESULTS FROM CLEO}

\author{ S.P. PAPPAS }

\address{California Institute of Technology, 
1200 East California Boulevard,\\
Pasadena CA, USA}

\maketitle\abstracts{
   The CLEO Collaboration has extracted improved values of \MVcb\ and
   \MVub\ from measurements of exclusive and inclusive decays of
   $B$~mesons. The measurement of \btodstlnu\ at zero recoil combined
   with the predicted form factor \Fdstwmin\ yields \MVcb. The photon
   energy spectrum in \btosg\ and the hadronic recoil mass spectrum in
   \btoxclnu\ determine non-perturbative HQET parameters used with
   inclusive \btoc\ and \btou\ rates to obtain \MVcb\ and \MVub.
}

\section{Introduction}

Decays of $B$~mesons provide a window into flavor physics, measuring
CKM matrix unitarity and \CP\ violation. The CKM matrix elements \Vcb\
and \Vub\ can be measured directly via $B$~meson decays (event yield
is proportional to \Vcb\ or \Vub), and the elements \Vts\ and \Vtd\
indirectly via $B$~mixing. At CLEO, $B$~mesons are produced almost at
rest, so mixing is not accessible (hence neither is \Vts\ nor
\Vtd). However, CLEO can measure \Vub\ and \Vcb\ well via
semi-leptonic decays of $B$~mesons. We present results from exclusive
and inclusive semi-leptonic branching fractions.

In exclusive decays we reconstruct final states and use form factors
from Heavy Quark Effective Theory (HQET)~\cite{HQET1,HQET2} to
extract the underlying CKM element.  In inclusive decays we assume
quark hadron duality, summing over the final states. Again, HQET
provides the decay dynamics but requires inputs accounting for
non-perturbative effects.

\section{The CLEO Experiment}

The CLEO experiment (described extensively
elsewhere~\cite{CLEO_I,CLEO_II,CLEO_III}) is located at Cornell
University in Ithaca, New York, USA. It is built around the CESR
$e^+e^-$ storage ring operating at a center of mass energy of
\nUnit{10.58}{\nGeV}.  Data is taken 2/3 at the \UpsIVS\ resonance and
1/3 continuum just below the resonance.  The \CLEOII\ (initial) and
\CLEOIIV\ (upgraded) detectors yielded respectively 1/3 and 2/3 of the
total \nUnit{13.5}{\ninvfb} luminosity used in these analyses.

\section{Exclusive \btodstlnu}

The \btodstlnu\ analysis measures the decay rate in the zero recoil
limit where HQET makes precise predictions for the \btodst\ form
factor $F_{D^*}(w)$ ($w = {v_B}^\mu\,{v_{D^*}}_\mu$, equivalent to
$q^2$). Accounting for phase space and spin physics ($\Phi(w)$), the
transition rate is:
\be
   \frac{d\Gamma}{dw} =
      \frac{G^2_F}{48\pi^2} |V_{cb}|^2 |F_{D^*}(w)|^2 \Phi(w)
   \label{eq:dstlnudec}
\ee

\begin{floatingfigure}[r]{0.37\textwidth}
\psfig{figure=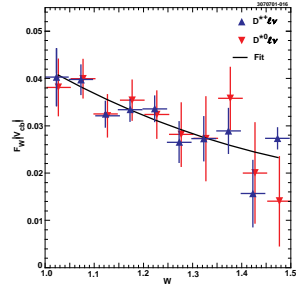,height=1.50in}
\caption{Measured $|V_{cb}|^2 |F_{D^*}(1)|^2$ as a function of $w$ for
charged and neutral modes of \btodstlnu. The curve is the fit
yielding $|V_{cb}|^2 |F_{D^*}(1)|^2$.
\label{fig:btodstlnu}}
\end{floatingfigure}

Zero recoil is at $w_{\rm min} = 1$ where the $D^*$ is stationary
relative to the $B$. For infinite heavy quark masses, the $b \ra c$
transition does not disturb the light quarks, so $F_{D^*}(w) \equiv
1$. For finite quark masses, HQET provides corrections:
$F_{D^*}(w_{\rm min}) = 0.913 \pm 0.042$.

The $D^*$ is reconstructed and combined with the lepton and the beam
energy to yield the decay missing mass in terms of the angle between
the $B$ and $D^*\ell$ momenta.  Assuming a missing neutrino, the decay
angle $\cos \theta_{B\,D^*\ell} \propto p_B\cdot ( p_{D^*} + p_\ell )$
can be calculated. For non-signal decays this will fall outside $[-1,
1]$, distinguishing them from signal.

We reconstruct charged and neutral modes, bin the yield in $w$ and
$\cos \theta_{B\,D^*\ell}$, and correct by $\Phi(w)$. This is fit to a
polynomial form with slope and curvature related by dispersion
relations~\cite{disp1,disp2} (Fig.~\ref{fig:btodstlnu}), yielding the
values~\cite{cleovcbx} $|V_{cb}|\,F_{D^*}(1) = (4.22 \pm 0.13 \pm
0.18)\times10^{-2}$ and $\rho^2 = 1.61 \pm 0.09 \pm 0.21$, and via
Eq.~\ref{eq:dstlnudec}:
\be
   |V_{cb}| = (46.4 \pm 1.4 \pm 2.0 \pm 2.1)\times10^{-3}
\ee

Work is also underway to reconstruct exclusive $b\ra u$ transitions,
relating \MVub\ to branching fractions for $B\ra \pi\,\ell\,\nu$ and
$B\ra (\rho/\omega)\,\ell\,\nu$.  Improved \MVub\ results are expected
by summer.

\section{\MVub\ from Lepton End Point}

\begin{floatingfigure}[l]{0.40\textwidth}
\vspace*{6mm}~\\
\hspace*{7mm}
\psfig{figure=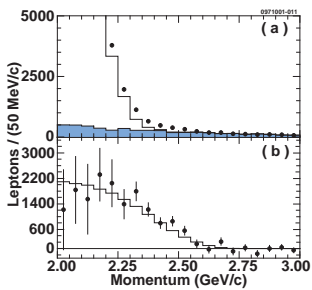,height=1.50in}\\
\vspace*{-4mm}
\caption{Measured lepton momentum spectra, total and background
subtracted in the end point region (\nUnit{2.2}{\nGeV} to
\nUnit{2.6}{\nGeV}) together with the fit to the predicted shape.
\label{fig:btoulepend}}
\end{floatingfigure}

The simplest measurement of \MVub\ counts events near the lepton
momentum end point, which can only be populated by \btou\
decays. Predicting the rate in the end point region is complicated by
the kinematics of the bound state $b$~quark, but these effects can be
related to \btosg\ decays via a shape function~\cite{shfn1,shfn2}.
Extrapolating from the end point to the total spectrum~\cite{lepxtrp}
yields \MVub.

We suppress continuum and other backgrounds by neural net and subtract
them, then measure the end point branching fraction: $\Delta{\cal
B}_{2.2\,-\,2.6\,{\rm GeV}} =$ $(2.30 \pm 0.15 \pm
0.35)\times10^{-4}$. The fraction of the total branching fraction in
this region is $f_u(p)_{2.2\,-\,2.6\,{\rm GeV}} = 0.130 \pm 0.024 \pm
0.015$, yielding a total rate ${\cal B}_{ub} = (1.77 \pm 0.29 \pm
0.38)\times10^{-3}$. We obtain~\cite{cleovublep}
(Fig.~\ref{fig:btoulepend}):
\be
   \begin{array}{rcl}
      |V_{ub}| & = & (3.06 \pm 0.08 \pm 0.08) \times10^{-3}\,\times\,
                     \sqrt{({\cal B}_{ub}/0.001)\cdot(1.6\,{\rm
                     ps}/\tau_B)} \\
               & = & (4.08 \pm 0.34 \pm 0.44 \pm 0.16 \pm 0.24)\times10^{-3}
   \end{array}
\ee
with uncertainties from $\Delta{\cal B}$, $f_u(p)$, $V_{ub}$ from
${\cal B}_{ub}$ (theory), and the shape function.

\section{Extracting non-perturbative HQET Parameters}

The inclusive semi-leptonic decay rate of $B$~mesons to charmed states
in HQET is~\cite{HQETVcb1,HQETVcb2}:
\be
   \Gamma(B \ra X_c\,\ell\,\nu) \propto
         \frac{G_F^2m_B^5}{192\pi^3}|V_{cb}|^2\left[
            1 + \left(\frac{\bar{\Lambda}}{m_B}\right)
              + \left(\frac{\bar{\Lambda},\lambda_1,\lambda_2}{m_B^2}\right)
              + {\cal O}\left(\frac{1}{m_B^3}\right)
         \right]
         + \mbox{rad. corr.}
   \label{eq:btoclnuincschem}
\ee
The parentheses represent functional forms depending on
non-perturbative quantities \Lbar\ ($m_B - m_b$), \lone\ (kinetic
energy of the bound $b$ quark), and \ltwo\ (hyperfine splitting,
$m_{B^*} - m_B$), and inverse powers of $m_B$. They cannot be
calculated ab initio, but are universal.  \Lbar\ and \lone can be
measured in \btosg\ and \btoclnu\ decays and applied via
Eq.~\ref{eq:btoclnuincschem} to extract \MVcb.

\subsection{\btosg\ Photon Energy Moments}

\begin{floatingfigure}[r]{0.60\textwidth}
\hspace*{\fill}\psfig{figure=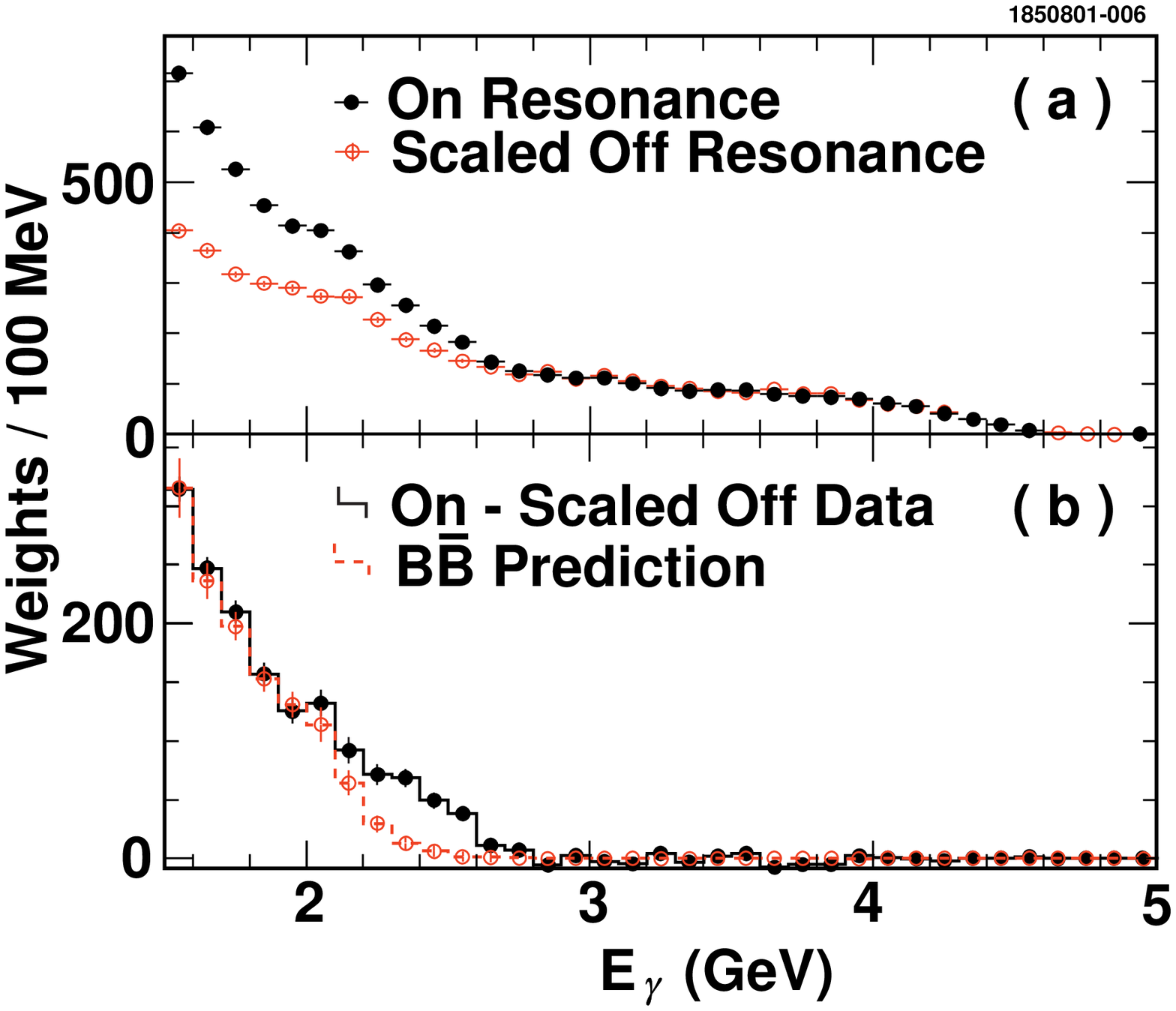,height=1.50in}~\psfig{figure=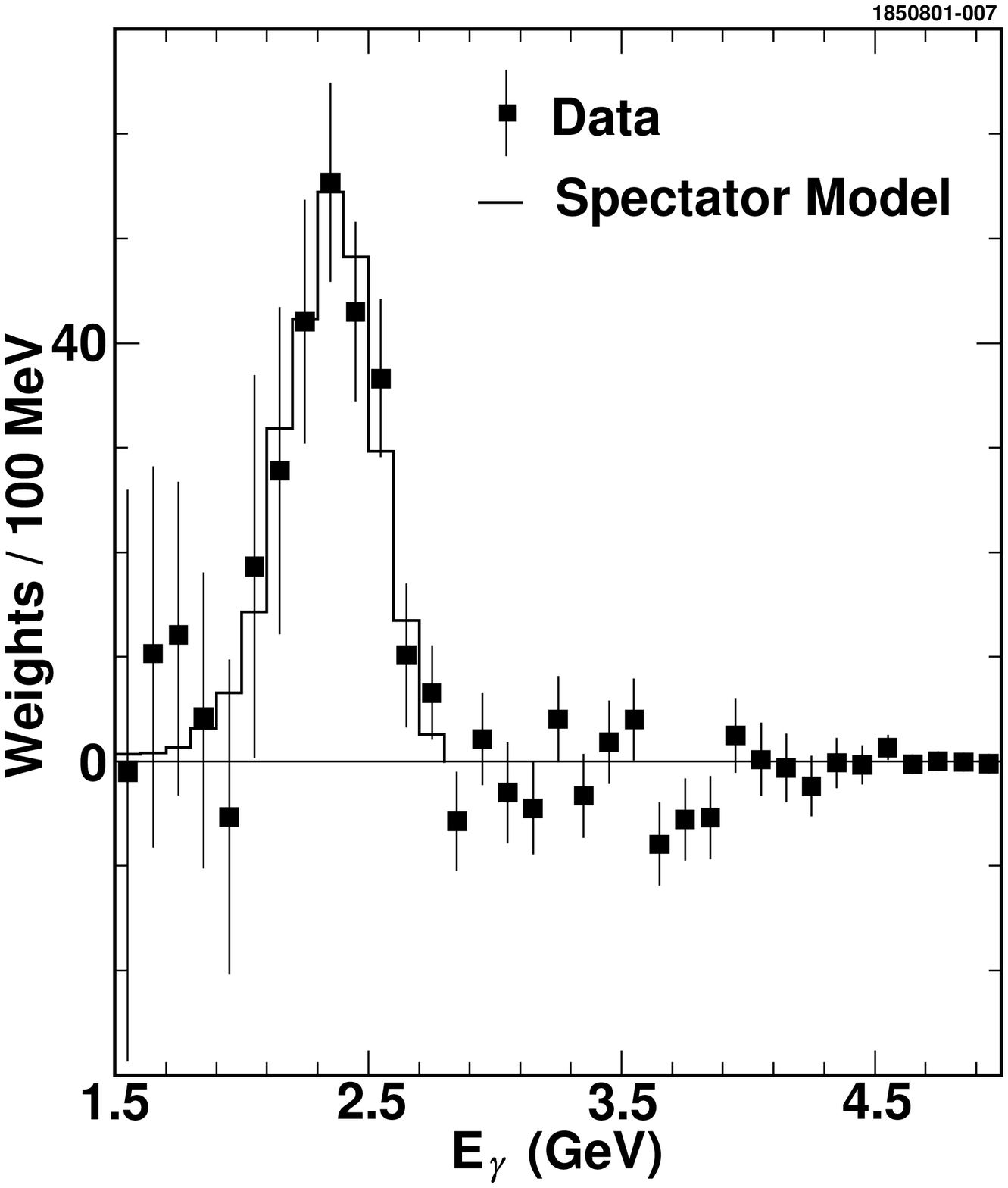,height=1.50in}\hspace*{\fill}
\caption{Plots of unsubtracted photon energy spectra in \btosg\ decays
and the inclusive spectrum after subtraction of all contributions
other than \btosg.
\label{fig:btosg}}
\end{floatingfigure}

Initially, \btosg\ decays were studied to seek non Standard Model
physics. Current branching fractions agree with
S.M. predictions~\cite{bsg1,bsg2,bsg3} and no \CP\ asymmetry is
observed, so the decay is now used to measure \Lbar. The moments
$\langle E_\gamma \rangle$ and $\langle E_\gamma^2 \rangle - \langle
E_\gamma \rangle^2$ of the inclusive \btosg\ photon spectrum are
naively $m_b/2$ and $E_{b,{\rm kin}}$, determining \Lbar\ and
\lone. However, only the expansion of $\langle E_\gamma \rangle$
converges.

We isolate the \btosg\ signature~\cite{cleobsg} by suppressing and
subtracting three orders of magnitude larger backgrounds (continuum
and \BBbar\ decays).  A neural net combines event shape, lepton
identification, and pseudo reconstruction into a signal probability
yielding the $\gamma$~spectrum (Fig.~\ref{fig:btosg}) with moments:
$\langle E_\gamma \rangle = 2.364 \pm 0.032 \pm 0.011 {\rm GeV}$ and
$\langle E_\gamma^2\rangle - \langle E_\gamma\rangle^2 = 0.0226 \pm
0.0066 \pm 0.0020 {\rm GeV}^2$.  HQET relates
these~\cite{HQETbsg1,HQETbsg2} to \Lbar\ ($M_H$ is $M_D$ or $M_B$):
\be
\langle E_{\gamma}\rangle =
            \frac{M_B}{2} [1 - 0.385 \frac{\alpha_s}{\pi} - 0.620 \beta_0
            (\frac{\alpha_s}{\pi})^2 - \frac{\bar \Lambda}{M_B} (1-0.954
            \frac{\alpha_s}{\pi} - 1.175 \beta_0
            (\frac{\alpha_s}{\pi})^2) + {\cal O}(1/M^3_H) ]
\ee

\subsection{\btoclnu Hadronic Mass Moments}

\begin{floatingfigure}[r]{0.35\textwidth}
\psfig{figure=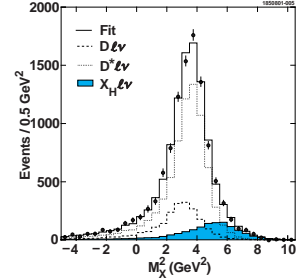,height=1.50in}
\caption{Hadronic recoil mass spectrum for \btoclnu\ decays, fit
to components for $D$, $D^*$, and heavier states.
\label{fig:hadmom1}}
\end{floatingfigure}

An HQET expansion~\cite{HQEThmm1,HQEThmm2} relates the moment
$\left<M^2_{Xc}\right>$ of the hadronic recoil mass in \btoclnu\
decays to \Lbar\ and \lone.  (The second moment, $\left<(M_X^2 -
M_D^2)^2\right>$, again does not converge). Measuring these
decays~\cite{hadmom_vcb} uses neutrino reconstruction techniques
pioneered by CLEO~\cite{nurec}.

Neutrino reconstruction relies on detector hermiticity and careful
modelling of energy flow. We sum the kinematics of all particles in
the event, and use event charge, lepton count, and the invariant mass
of the inferred neutrino to ensure the measurement precisely reflects
the kinematics of the semi-leptonic decay.  The hadronic recoil system
is calculated from the $B$, $\ell$, and $\nu$ kinematics, neglecting
the (small) term $\vec{p}_B\cdot\vec{p}_{\ell\nu}$.

\begin{floatingfigure}[r]{0.35\textwidth}
\vspace*{-3mm}
\psfig{figure=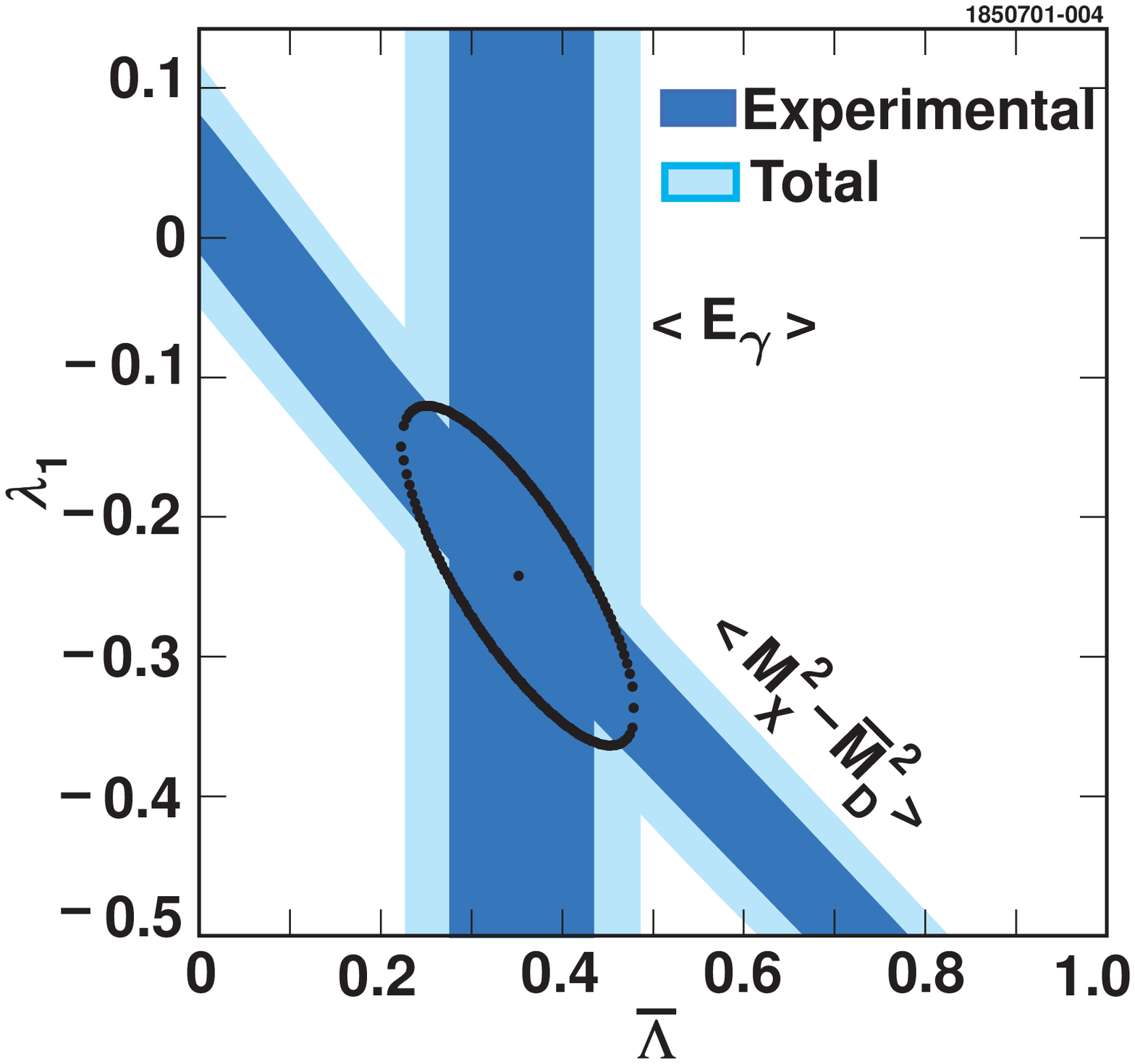,height=1.50in}
\vspace*{-4mm}
\caption{The simultaneous fit of \Lbar\ and \lone\ from \btosg\ and
\btoclnu.
\label{fig:hadmom2}}
\end{floatingfigure}

The recoil mass spectrum is fit to three components
(Fig.~\ref{fig:hadmom1}): \btodlnu, \btodstlnu, and \btoxhlnu
($D^{**}$, etc. and non-resonant). From this we
measure~\cite{hadmom_vcb} $\left<M_x^2 - M_D^2\right> = 0.251 \pm
0.023 \pm 0.062 {\rm GeV}^2$ and $\left<(M_X^2 - M_D^2)^2\right> =
0.639 \pm 0.056 \pm 0.178 {\rm GeV}^4$.

Combining the results of \btosg\ and \btoclnu\ measurements we obtain
simultaneous constraints on \Lbar\ and \lone. We perform a fit
(Fig.~\ref{fig:hadmom2}), yielding~\cite{hadmom_vcb}:
\be
   \begin{array}{rcl}
      \bar{\Lambda} & = & 0.35 \pm 0.07 \pm 0.10 {\rm GeV} \\
      \lambda_1 & = & -0.238 \pm 0.071 \pm 0.078 {\rm GeV}^2
   \end{array}
   \label{Eq:hqetpars}
\ee

\section{\MVcb\ from Inclusive Semi-Leptonic $B$~Decays}

Recalling Eq.~\ref{eq:btoclnuincschem}, we can relate the rate of \btoc\
semi-leptonic decays to \MVcb.
The rate has been measured via a two lepton tag
technique~\cite{cleobestbtoc} and is corrected for \btou\ to extract
${\cal B}(B \ra X_c\,\ell\,\nu) = (10.39 \pm 0.46)\%$.  Using the
measured admixture fraction~\cite{cleobestchgrat} ($f_{+-}/f_{00} =
1.04 \pm 0.08$) and the lifetimes ($\tau_{B^\pm} = (1.548 \pm
0.032)\,{\rm ps}$ and $\tau_{B^0} = (1.653 \pm 0.028)\,{\rm ps}$) we
determine $\Gamma_{\rm sl} = (0.427 \pm 0.020)\times10^{-10} {\rm
MeV}$, finally yielding \MVcb\ via measured \Lbar, \lone, and \ltwo:
\be
   V_{cb} = (40.4 \pm 0.9 \pm 0.5 \pm 0.8)\times10^{-3} = (40.4 \pm
      1.3)\times10^{-3}
\ee
This result has an error of only $3.2\%$, making it the most precise
determination of \MVcb\ to date.

\section*{Acknowledgments}

The author would like to thank the researchers at CLEO and CESR for
providing such remarkable data and analyses as well as the agencies
DOE and NSF for funding this research. Thanks also to the organizers
of the XXXVII Recontres de Moriond for a wonderful conference and
environment to study and discuss physics.

\end{document}